\documentstyle[12pt,twoside,fleqn,espcrc1]{article}

\newcommand{\AmS}{{\protect\the\textfont2
  A\kern-.1667em\lower.5ex\hbox{M}\kern-.125emS}}
\title{
Open Charm Production in Hadronic and Heavy-Ion
Collisions at
RHIC and LHC Energies to $O(\alpha_s^3)$}
\author{Ina Sarcevic\address
{Department of Physics,
        University of Arizona, \\
        Tucson, AZ 85721}}

\begin{document}
\maketitle

\begin{abstract}
We present results on rapidity and transverse momentum
distributions of inclusive charm quark production in
hadronic and heavy-ion collisions at RHIC and LHC energies,
including the next-to-leading
order, $O(\alpha_s^3)$, radiative corrections and the
nuclear shadowing effect.
We determine the hadronic and the {\it effective} (in-medium)
K-factor
for the differential and total inclusive charm cross sections.
We show that these
K-factors have strong $p_T$ dependence.
We discuss how measurements of charm production at RHIC and LHC
can provide valuable information about the
gluon density
in a nucleus.

\end{abstract}

\section{INTRODUCTION}

Recently, there has been a considerable theoretical and experimental
interest in studying heavy-quark production in hadronic and
nuclear collisions.  Theoretical calculation of the
heavy-quark differential and total cross sections has been
improved by including the next-to-leading
order, $O(\alpha_s^3)$,
radiative corrections \cite{NAS89,MAN92}.  For bottom and charm production
these corrections are large, especially at threshold energies and
at very high energies.
Future high-precision measurements of the charm
production at Fermilab fixed-target experiments
could therefore provide a
stringent test of perturbative QCD \cite{CHM2000}.
In addition,
studying charm production at the proposed
heavy-ion colliders, BNL's Relativistic Heavy Ion
Collider (RHIC) and CERN's Large Hadron Collider (LHC) is of
special interest.
Open charm production has
been suggested as an elegant signal for detecting the
formation of quark gluon plasma (QGP) in heavy-ion collisions
\cite{SHU92}.
If thermalization is reached
in heavy-ion
collisions at RHIC
and LHC,
we expect a very dense matter, of the order of
few
GeV/$fm^{3}$, to be formed in the
initial stage of the collision \cite{Mu}.  It seems plausible that this
density is sufficient for creating a new
state of matter, the QGP.
The
search for a clean, detectable signal for this new state of matter
has been one of the most challenging
theoretical problem for the last few years.  Some of the
proposed signals, thermal photons, dileptons and $J/\Psi$
suppression, have been studied and found to be difficult to detect due
to the large QCD
background \cite{Vesa}.
In
order that the enhanced charm production can be used as the signal of
QGP, we need to understand the
QCD background, namely
the production of charm quarks through the
hard collisions of partons inside the nuclei.
This type of charm production
at RHIC and LHC energies is dominated
by initial-state gluons.  Therefore, in addition to the
possibility of
pointing towards the formation of quark-gluon plasma in
high-energy heavy-ion collisions,
combined measurements of charm
production in p-p, p-A and A-A collisions could
provide valuable information about the
gluon density in a nucleus.
\section{CHARM PRODUCTION IN HIGH ENERGY
HADRONIC AND NUCLEAR COLLISIONS}
We calculate the total
cross section for charm production in hadronic
collisions by convolving the
parton densities in the hadron with the
hard-scattering parton cross sections.
Our calculation
includes both the leading order subprocesses, $O(\alpha_s^2)$, such as
$ q+\bar q\rightarrow Q+\bar Q$ and
$ g+g\rightarrow Q+\bar Q$,
and next-to-leading order subprocesses, $O(\alpha_s^3)$, such as
$ q+\bar q\rightarrow Q+\bar Q +g$,
$ g+q \rightarrow Q+\bar Q+g$,
$ g+\bar q \rightarrow Q+\bar Q+\bar q$ and
$ g+g \rightarrow Q+\bar Q+g $.
Due to the fact that
the hadronic structure functions have not been measured below
$Q^2 = 8.5 GeV^2$ \cite{WOL94},
we take factorization scale
to be $2m_c$.  We
consider renormalization
scale between $m_c$ and $2m_c$.
In our calculation we use two-loop-evolved
parton structure functions,
MRS A
\cite{MAR95},
as a canonical set.
In order to reduce theoretical uncertainty due to the choice of the
renormalization scale,
we
compare our calculation of the total cross section
to the current
data for charm production in pp and pA
collisions for beam energies ranging
from $50 GeV$ to $2 TeV$ \cite{AOK89}.  We find
better agreement with the data when we use the
renormalization scale $Q^2=m_c^2$.  At these energies,
the uncertainty due to the
choice of the structure function is negligible.
\subsection{The Nuclear Shadowing Effect}
To determine charm production in nuclear
collisions  we need to know the parton
densities {\it in-medium} and the geometrical (spatial)
overlapping function.
If nucleons were independent,
the parton densities in a nucleus would
simply be A times the parton density in a nucleon.  At high
energies,  the parton densities become so large that the sea quarks
and gluons from different nucleons overlap spatially and the nucleus
cannot be treated as a collection of uncorrelated nucleons.  This
interactions of partons through
annihilation effectively reduce the parton densities in a
nucleus.  Recent EMC,
NMC, and E665 measurements of $F_2^A/F_2^D$
in the small-$x$ region have confirmed this picture \cite{NMC}.
Furthermore, these measurements indicate that shadowing effect
has strong $x$-dependence.
The next-to-leading order subprocesses
also have different contributions in different regions of
phase space.  Thus, we also calculate the
single inclusive distributions, such as the rapidity and transverse
momentum distributions \cite{SAR94}.
We obtain the
overlapping function by integrating the
Woods-Saxon
nuclear densities
over the appropriate overlapping spatial region.  In case of
Au-Au collisions, the value for
this function at small impact parameter (i.e. for
central collisions) is $30.6mb^{-1}$.
\subsection{Rapidity and Transverse Momentum Distributions for
Charm Production at RHIC and LHC Energies}

In Fig. 1 we present our results for
the rapidity distributions of charm quarks produced in
hadronic (short-dashed line) and Au-Au collisions (solid line)
at RHIC and LHC energies.  We include the next-to-leading
order corrections, $O(\alpha_s^3)$, and the nuclear
shadowing effect \cite{SAR95}.  From Fig.
1 we note that
the hadronic K-factor is $2$ at RHIC and $2.2$ at LHC in the
central rapidity region.
To illustrate the importance of the $O(\alpha_s^3)$
corrections
{\it combined} with nuclear shadowing effects
in different regions of phase space we
determine the
{\it effective} (i.e. in-medium)
K-factor, defined as the ratio of the inclusive distribution for
charm production in A-A collisions to the
leading-order distribution {\it without} nuclear effects.
We find the effective K-factor to be $1.5$ at RHIC and $1.3$ at
LHC in the central rapidity region.
In Fig. 2 we present the $p_T$-distribution of charm quark
produced in hadronic (short-dashed line) and Au-Au collisions
(solid line).  We find that
both hadronic and in-medium K-factor have strong
$p_T$ dependence,
increasing from $1.7$ ($1.1$) at $p_T=1$GeV  to
$3.1$ ($3$) at $p_T=6$GeV at RHIC and from
$1.5$ ($0.7$) at $p_T=1$GeV to $7.6$ ($4.6$) at
$p_T=6$GeV at the LHC.
The behavior of
the {\it effective} K-factor is a direct consequence of the
fact that the low $p_T$ region (or small $x$) corresponds to
the maximum shadowing of the gluon distribution, which in
case of gold is about $62\%$, while the larger values of
$p_T$ probe region of phase space where the nuclear shadowing
is smaller.

\vskip 3 true in
\noindent
Figure 1.  The rapidity distribution for charm quark production
at RHIC and LHC.
\vskip 3 true in
\noindent
Figure 2.  The transverse momentum distribution
for charm quark production
at RHIC and LHC energies.

\vskip 0.2 true in

Theoretical uncertainty in the calculation of
charm production in nuclear collisions due to the choice of
the structure function is small at Fermilab fixed target
energies (only few percent) and at RHIC ($8\%$
in the central rapidity region
), while at LHC
energies this uncertainty is about a factor of $6$.
Furthermore, the shape of the rapidity distribution and
the transverse momentum distribution at LHC is
very sensitive to the small-$x$ behavior of the gluon
structure function.
We find that the dominant contribution
to charm production
comes from the initial state gluons
(about $95\%$ at RHIC and
$99\%$ at LHC).
Thus,
combined measurements of inclusive charm production in hadronic and
nuclear collisions at these energies,
in addition to providing an important test of perturbative QCD in
the small-$Q^2$ and small-$x$ region,
might be able to
provide
valuable information about the
elusive role of gluons
inside a nucleus, especially
in the region of very small $x$.

\section{ACKNOWLEDGEMENTS}
The results presented here were obtained
in collaboration with P. Valerio.
This work was supported in part through
U.S. Department of Energy Grant
DE-FG03-93ER40792.

\end{document}